\documentclass[aps,prl,twocolumn,showpacs,10pt,superscriptaddress,preprintnumbers,nofootinbib,longbibliography]{revtex4-1}
\usepackage{epsfig,amssymb,amsmath,psfrag,epstopdf,color,siunitx}
\pdfoutput=1
\usepackage{graphicx}
\usepackage[caption=false]{subfig}
 \usepackage[export]{adjustbox}

\usepackage{paralist}
\usepackage{hyperref}
\usepackage{upgreek}
\allowdisplaybreaks

\newcommand{\nn}{\nonumber}

\definecolor{orange}{rgb}{0.9, 0.25, 0}
\definecolor{pur}{rgb}{0.6,0.,1.}

\DeclareRobustCommand{\Fig}[1]{Fig.~\ref{#1}}

\AtBeginDocument{\DeclareRobustCommand{\Ref}[1]{Ref.~\cite{#1}}}

\newcommand{\eq}[1]{Eq.~\eqref{eq:#1}}
\newcommand{\eqs}[2]{Eqs.~\eqref{eq:#1} and \eqref{eq:#2}}
\newcommand{\fig}[1]{Fig.~\ref{fig:#1}}

\newcommand{\Afb}[1]{A_{\mathrm{FB}}^{#1}}

\begin{document}
\title{Flavor Flux Correlators, Forward-Backward Asymmetries, and Anomalies}

\author{Kyle Lee}
\email{kyle@anl.gov}
\affiliation{High Energy Physics Division, Argonne National Laboratory, Lemont, IL, USA}

\author{Ian Moult}
\email{ian.moult@yale.edu}
\affiliation{Department of Physics, Yale University, New Haven, CT 06511, USA}

\begin{abstract}
Flavor plays a crucial role in the structure of the Standard Model. As such, many important collider physics measurements incorporate the flavor information of detected hadrons.
Recent progress in the study of energy correlators has enabled collider physics measurements of energy flux to be reformulated as operator statements.
In this \emph{Letter}, we introduce flavor flux correlators, which measure the angular distribution and correlations of flavor quantum numbers.
We emphasize a unique feature of heavy flavor flux correlators as compared to correlators of other charges, namely the $b$ and $c$ flavor quantum numbers are gapped at the heavy quark mass scale, making them a particularly clean perturbative observables, with $\mathcal{O}(\Lambda_{\text{QCD}}/m_Q)$ suppressed non-perturbative power corrections.
As an application, we reformulate the heavy quark forward-backward asymmetry as a macroscopic consequence of a mixed electroweak anomaly of the heavy-quark flavor current.
Using the relation to the anomaly, we compute the flavor flux correlator analytically at next-to-next-to-next-to-leading order and verify using parton shower simulations that non-perturbative corrections are suppressed, as expected from the gapped nature of the heavy flavor quantum numbers.
Flavor flux correlators can be measured using archival electron-positron collision data, as well as at Belle~II and future colliders. 
\end{abstract}

\maketitle

\section{Introduction}

The electroweak sector of the Standard Model exhibits a rich flavor structure.
Collider physics measurements, which have carefully unveiled this structure with ever increasing precision, rely crucially on the incorporation of flavor information.
This can be achieved either at the level of individual detected hadrons, as in flavor-tagged measurements of neutral-meson oscillations, or through the probabilistic association of jet-level information with underlying partonic degrees of freedom, as often incorporated in new physics searches.
As compared to energy flux, the distribution of flavor fluxes produced in a collision typically depends sensitively on the non-perturbative physics of the hadronization process, making the incorporation of flavor information challenging for precision theoretical calculations.

Famous examples of Standard Model measurements which necessitate the precision theoretical description of flavor fluxes are the heavy quark ($b$ and $c$) forward-backward asymmetries in electron-positron collisions at the $Z$-pole. 
Due to parity violation in the electroweak sector, the structure of the $Zbb$ ($Zcc$) vertex produces \emph{b(c)-quarks} with an asymmetric angular distribution, which is ultimately imprinted as an asymmetry in the distribution of \emph{b(c)-mesons} reconstructed experimentally. 
However, due to confinement, achieving precision measurements of the $Zbb(Zcc)$ vertex structure from the asymptotic fluxes of hadrons is non-trivial.
Traditional measurements of $\Afb{b,c}$ proceed by assigning to each event an underlying quark direction and a charge constructed from jets or hemispheres~\cite{OPAL:1993wua,L3:1992fsb,ALEPH:2001mdb,DELPHI:2003fml,DELPHI:2004wzo,OPAL:2003pfe,DELPHI:1991mqi,OPAL:1992jsm,ALEPH:1996qlh,ALEPH:1998pmr}. Such an assignment receives $\mathcal{O}(\alpha_s)$ corrections with $\mathcal{O}(1)$ coefficients that depend on the choice of the axis (thrust or jet axes) and of the charge estimator (jet charge, vertex charge, lepton tags), as well as on the cuts~\cite{Djouadi:1994wt,Altarelli:1992fs,Ravindran:1998jw,Catani:1999nf,Weinzierl:2006yt,Bernreuther:2016ccf,Bernreuther:2023jgp,Wang:2020ell} and hadronization corrections must be modeled. For a modern discussion of the QCD uncertainties associated to these measurements see~\cite{dEnterria:2018jsx,AlcarazMaestre:2020fmp}. Improving the theoretical description of heavy quark asymmetries is not merely an academic exercise: the b-quark forward-backward asymmetry is currently the largest tension in global electroweak fits \cite{Baak:2014ora,Haller:2018nnx,Fischer:2026bka}, and is quite natural in many extensions of the Standard Model which couple preferentially to the third generation \cite{Haber:1999zh,Peccei:1990uv,Choudhury:2001hs,Morrissey:2003sc,Kumar:2010vx,Batell:2012ca,He:2002ha,Djouadi:2006rk,DaRold:2010as,Liu:2017xmc,Glioti:2024hye,Allwicher:2023shc,Davighi:2023iks}. Additionally, future colliders will significantly reduce the experimental uncertainties, necessitating improved theoretical techniques.

A new way of thinking about measurements of fluxes at colliders was presented in \cite{Hofman:2008ar}, in the simplified context of conformal field theories. Measurements of fluxes can be expressed as correlation functions of \emph{detector operators} formed from the currents of the charges being measured, providing a sharp connection between collider physics measurements and correlation functions of currents. This perspective has been highly successful for energy flux \cite{Moult:2025nhu}. Recently, it was shown~\cite{Zhang:2026emt} that the forward-backward asymmetry of electromagnetic charge on all hadrons can be formulated in terms of a correlator of electromagnetic charge, providing a beautiful interpretation of forward-backward asymmetries at the $Z$-pole as macroscopic consequences of mixed electroweak anomalies. More excitingly, using archival DELPHI data, it was shown that this can be achieved in practice~\cite{Zhang:2026vdo,Zhang:2026esd}, providing a first measurement of a charge correlator, and showing the experimental feasibility of a new approach to studying forward-backward asymmetries.

Building on these successes, in this \emph{Letter}, we introduce \emph{flavor flux correlators}, allowing a broad generalization of the energy correlator program to correlators incorporating flavor information.
While for light flavors, $f=u,d,s$, flavor flux correlators are sensitive to non-perturbative hadronization effects, we point out that the phenomenologically most interesting fluxes, those of $b$ and $c$, are protected. The mass gap $m_Q\gg\Lambda_{\rm QCD}$ removes the heavy flavor quantum number from the low energy theory, leaving parametrically suppressed hadronization corrections of order $\Lambda_{\rm QCD}/m_Q$. As an application of flavor flux correlators, we reformulate the heavy quark forward-backward asymmetry as a macroscopic consequence of a mixed electroweak anomaly of the heavy-quark flavor current. Parton shower simulations confirm the expected suppression of the non-perturbative corrections from the gapped nature of the heavy flavor quantum numbers, and we perform a precision calculation of the asymmetry.

\begin{figure*}[t]
\centering
\includegraphics[width=\textwidth]{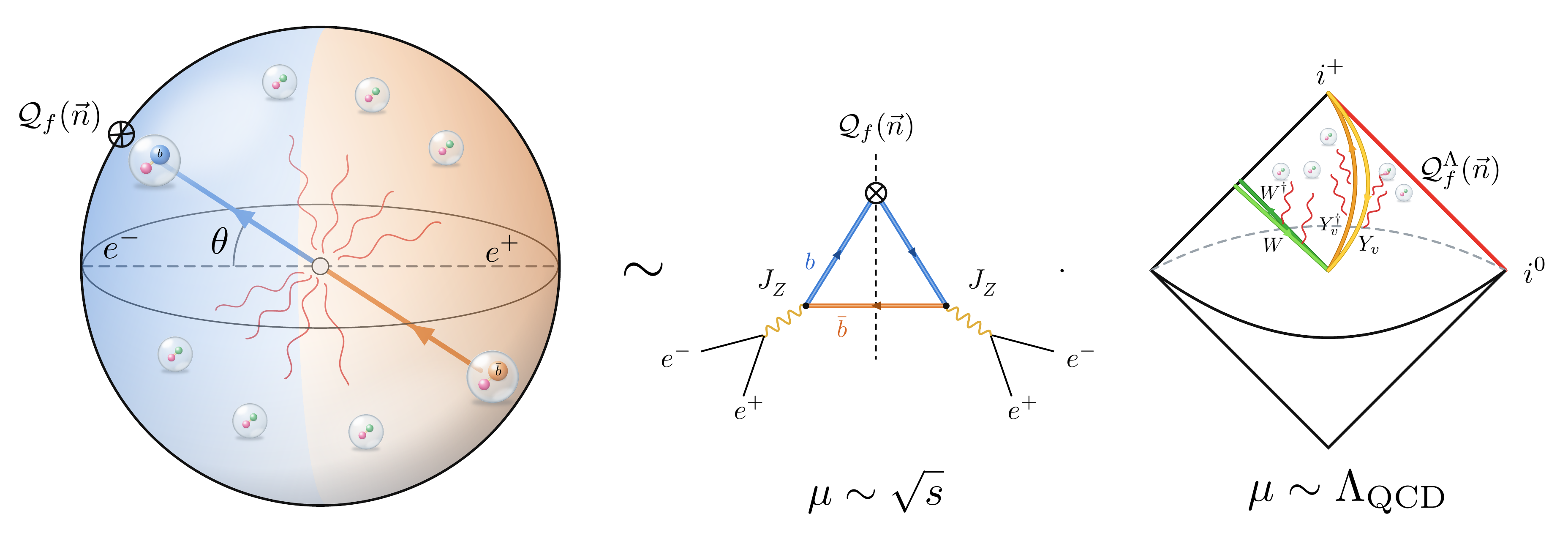}
\caption{Factorization of the heavy-flavor flux measured on hadrons (left) into a short-distance contribution (center) and a low-energy detector matrix element (right). The short-distance coefficient $C_A^{\rm NS}$ is related to the mixed electroweak anomaly. Below the heavy-quark mass, interactions of the heavy source with light degrees of freedom are encoded in the Wilson lines $Y_v$ and $W$. Since these light degrees of freedom carry no heavy flavor, the low-energy detector $\mathcal{Q}_{f_Q}^{\Lambda}$ is transparent.}
\label{fig:flavorflow}
\end{figure*}

\section{Flavor Flux Correlators}
\label{sec:formalism}

The measurement of asymptotic fluxes of energy was formulated in terms of detector operators by Korchemsky and Sterman \cite{Korchemsky:1997sy}. This philosophy was extended in \cite{Hofman:2008ar}, where the measurement of the flux of a U(1) charge with current $J_F^\mu$ was expressed in terms of the detector operator
\begin{align}
\label{eq:detector}
\mathcal{Q}_F(\vec{n}) = \lim_{r\to\infty} r^2\int_0^\infty d t\; n_i\, J_F^i(t, r\vec{n})\,.
\end{align}
In a theory with asymptotic states, such as QCD, the action of the charge detector on asymptotic states simply weights each hadron by its quantum number,
\begin{align}
\label{eq:detector_action}
\mathcal{Q}_F(\vec{n})\,|X\rangle = \sum_{h\in X} F(h)\, \delta^{2}(\vec{n}-\vec{n}_h)\,|X\rangle\,.
\end{align}
Correlators of charge detectors, $\langle \mathcal{Q}_F(\vec{n}_1) \cdots \mathcal{Q}_F(\vec{n}_N) \rangle$, thus provide a direct relation between observables which can be directly measured from hadrons observed in detectors, and correlation functions of currents. Correlators of a variety of different charges, such as R-charges, electromagnetic charge, baryon number, have been studied in a number of different theories \cite{Hofman:2008ar,Belitsky:2013bja,Belitsky:2013xxa,Cordova:2017zej,Chicherin:2020azt,Riembau:2024tom,Cao:2026fzq,Monni:2025zyv,Zhang:2026emt}. Most recently, it has been shown that such correlators can actually be measured in real-world experiments, with the first measurement of the one-point correlator of electromagnetic charge in hadronic collisions at the $Z$-pole \cite{Zhang:2026emt,Zhang:2026esd,Zhang:2026vdo}.

Neglecting Yukawa interactions, the Standard Model exhibits a rich set of global flavor symmetries (see e.g. \cite{DAmbrosio:2002vsn}) from which we can form charge detector operators. While these symmetries are violated by electroweak interactions, we are mainly interested in the case of electron-positron collisions near the $Z$-pole, where electroweak bosons are not produced in the collision and act only as background fields. At such energies the effect of electroweak interactions is to cause the decays of heavy flavor mesons, as well as neutral meson oscillations. Heavy flavor meson decays occur at timescales set by the electroweak interactions, allowing the kinematics of mesons to be experimentally reconstructed, as if they were stable asymptotic states. Neutral meson oscillations are a small unavoidable effect that arises in all asymptotic measurements of flavor. We will return to the practical implications of including neutral-meson mixing in the End Matter.

We are therefore led to extend the study of detector correlators to consider flavor flux correlators, $\mathcal{Q}_f$, $f=u,d,s,c,b$, associated with the vector currents $J_f^\mu = \bar{f}\gamma^\mu f$, for which $f(h)$ counts the net number of valence $f$ quarks in $h$. For example, $b(B^-)=b(\bar{B}^0)=+1$ and $b(B^+)=b(B^0)=-1$. This provides a new set of correlators $\langle\mathcal{Q}_f(\vec{n}_1)\cdots \mathcal{Q}_f(\vec{n}_N)\rangle$. More generally, one can consider mixed correlators between different flavors, or with energy or charge detectors. 

At the $Z$-pole, flavor flux correlators also take a simple theoretical structure. The state produced in electron-positron collisions tuned to the $Z$-pole can be viewed as the action of the $Z$-boson current $J_Z^\nu$ acting on the vacuum. Correlators of the flavor flux in $e^+e^-$ annihilation at the $Z$-pole, constructed from an arbitrary number of flavor detectors placed at directions $\vec{n}_1,\ldots,\vec{n}_N$ on the celestial sphere, are given by
\begin{align}
\label{eq:correlator_defN}
&\langle\mathcal{Q}_f(\vec{n}_1)\cdots \mathcal{Q}_f(\vec{n}_N)\rangle \equiv \nn\\
&\quad\frac{L_{\mu \nu} \int d^4 x\, e^{i q \cdot x} \langle 0| J_Z^{\mu \dagger}(x)\, \textstyle\prod_{i=1}^{N}\mathcal{Q}_{f}(\vec{n}_i)\, J_Z^\nu(0)|0\rangle}{L_{\mu \nu} \int d^4 x\, e^{i q \cdot x} \langle 0| J_Z^{\mu \dagger}(x)\, J_Z^\nu(0)|0\rangle}\,,
\end{align}
where $q^2=s$ with $\sqrt{s}$ the CM energy, $L_{\mu\nu}$ is the leptonic tensor of the unpolarized $e^+e^-$ beams, and $J_Z^\mu$ is the current coupling to the $Z$. These objects are the flavor analogs of the multi-point energy correlators, and characterize the complete angular distribution of flavor in the final state. Their relation to correlators of $N+2$ currents will allow particularly interesting relationships to field theory phenomena, such as anomalies.

\section{Detector Matching for Gapped Quantum Numbers}
\label{sec:sumrules}

Detector operators are defined in the infrared, and therefore in QCD, they act on \emph{hadronic} asymptotic states. The non-perturbative nature of confinement makes a precise description of detector correlators challenging, and is a major obstacle to their use in precision measurements.
For the specific case of energy detectors, non-perturbative corrections are suppressed by $\mathcal{O}(\Lambda_{\text{QCD}}/\sqrt{s})$.
For more general charges, such as baryon number or electric charge, the situation is more subtle. 
This subtlety arises due to the fact that the perturbative Hilbert space of asymptotic states includes states with fractional baryon number or charge, while the non-perturbative Hilbert space of asymptotic states consists of confined hadrons with integer baryon number or charge. This necessitates a non-perturbative theoretical description. 

Modern descriptions rely on factorization theorems and effective field theories to isolate universal non-perturbative matrix elements. For the particularly simple case of a one-point charge correlator, this leads to the factorization
\begin{align}
\label{eq:factorization}
\hspace{-0.45cm}\frac{d\Sigma_{f}}{d \cos \theta} \equiv 2\pi\,\sigma_{\rm tot}\, \langle\mathcal{Q}_f(\vec{n})\rangle = \frac{3}{2}\,\cos \theta \; C_{A}^{\rm NS} \sum_q A_q^{(0)}\, \mathcal{M}_q^{f}\,,
\end{align}
where $\theta$ is the angle of the detector to the incoming electron, $\sigma_{\rm tot}$ is the total hadronic cross section, and azimuthal symmetry has been used to integrate over the detector azimuth. The factorization holds at leading power in $\Lambda_{\rm QCD}/\sqrt{s}$. The coefficient $A_q^{(0)}$ is the Born-level electroweak asymmetric normalization for quark flavor $q$, with the sum running over the five quark flavors and antiquark contributions included through charge conjugation, and $C_{A}^{\rm NS}$ is the first moment of the asymmetric non-singlet coefficient function, given in \eq{CANS} below. Finally, the nonperturbative input is isolated in the moments
\begin{align}
\label{eq:moment_def}
\hspace{-0.25cm}\mathcal{M}_q^{f} \equiv \sum_h \frac{f(h)}{2}\, \big[D_{q\to h}(N{=}1,\mu)-D_{\bar q\to h}(N{=}1,\mu)\big]\,,
\end{align}
of the fragmentation functions $D_{i\to h}(z,\mu)$, particular matrix elements in the low energy theory, defined in terms of matrix elements involving Wilson lines.

It is well known, although not often emphasized in discussions of QCD factorization, that the Hilbert space in the presence of defects, referred to as the defect Hilbert space, is different from the Hilbert space without defects. A famous illustration of the modification of the theory in the presence of defects is the Anderson orthogonality catastrophe \cite{Anderson:1967zze}. In the case of QCD, in the presence of a fundamental Wilson line, the defect Hilbert space houses excitations with fractional baryon and electric charge. In perturbation theory, the defect Hilbert space and the Hilbert space of asymptotic states are the same, but non-perturbatively they are different. 

We emphasize this point, since it plays a crucial role in understanding the flux of conserved charges, such as electromagnetic charge, or light flavors, in QCD. For conserved charges, it is often stated that one has a sum rule
\begin{align}
\label{eq:sumrule}
\sum_h f(h)\, D_{q\to h}(N{=}1,\mu) = f(q)\,,
\end{align}
giving $\mathcal{M}_q^f = f(q) = \delta_{qf}-\delta_{\bar q f}$, and eliminating the non-perturbative fragmentation function from the description of charge flux correlators in QCD. However, fragmentation functions are defined as matrix elements involving Wilson lines, and this sum rule arises from the \emph{incorrect} identification of sums over states in the defect Hilbert space, and the Hilbert space of asymptotic states in QCD. These Hilbert spaces can be identified in perturbation theory, so that this sum rule holds to every order in perturbation theory, but it does not hold non-perturbatively. This implies that correlators of conserved fluxes in QCD can receive non-perturbative corrections that are not suppressed by powers of $\Lambda_{\text{QCD}}/Q$.\footnote{Note that the momentum sum rule, by contrast, is protected. The soft hadrons responsible for the violations carry momentum fractions $x\sim \Lambda_{\rm QCD}/Q$, and hence a parametrically small fraction of the energy flux, whereas charge and flavor weight every hadron equally, independently of its momentum fraction.} This was emphasized and clearly explained using a different language in \cite{Collins:2023cuo,Kotlorz:2025xso}, and more recently a first attempt at characterizing these non-perturbative effects was performed in \cite{Ke:2026zia}. There has been significant recent progress in the formal understanding of defects in quantum field theory \cite{Billo:2016cpy}, including null defects \cite{Erramilli:2025pfh,Cuomo:2026mop}. It would be interesting to use these advances to improve the understanding of charge correlators, or to relate these non-perturbative corrections to the screening of Wilson lines \cite{Aharony:2023amq}, but for now these non-perturbative corrections remain a difficulty in the description of correlators of ungapped charges.

This same difficulty arises for flavor flux correlators of light flavors. However, a main point of this \emph{Letter} is to point out that the case of heavy flavor flux (e.g. $b$, $c$) is unique amongst the charges in the Standard Model in that it is conserved, and gapped at a scale $m_Q \gg \Lambda_{\text{QCD}}$. We will show that in this case, non-perturbative corrections are indeed suppressed by $\Lambda_{\text{QCD}}/m_Q$. Heavy flavor flux thus provides the first example of a completely computable charge correlator in the Standard Model, and most excitingly, it is the case of heavy flavor which is most important experimentally, opening the door to the use of flavor flux correlators for precision physics.

In the above language, the distinction between correlators of heavy flavor fluxes and correlators of light conserved charges is particularly clear: in a process involving heavy quarks, we can integrate out the dynamics at the scale $m_Q$, leaving a network of defects in the low energy theory \cite{Korchemsky:1991zp}. This is illustrated for the one-point correlator, in the case that a single heavy quark is produced, in \Fig{fig:flavorflow}. Crucially, for detectors of heavy flavor flux, the entire quantum number is integrated out at the scale $m_Q$, and does not exist in the low energy theory. In other words, the detector in the low energy theory becomes transparent. This protects heavy flavor flux correlators from non-perturbative contributions, and makes them computable entirely in perturbation theory up to $\Lambda_{\text{QCD}}/m_Q$ suppressed corrections.

This argument can be made rigorous in the language of Heavy Quark Effective Theory (HQET). 
The hadron-level detector $\mathcal{Q}_{f_Q}$ can be first matched onto HQET detectors with matching coefficients given by the HQET matrix elements $\chi_Q^{H_Q}$, expressed in terms of Wilson lines dressed by the light degrees of freedom, which absorb all infrared scales below $m_Q$. The HQET detectors can further be matched onto the partonic detectors according to \eq{factorization}, for which $m_Q$ would then be the infrared scale and $\sqrt{s}$ the ultraviolet one. Physically, this two-stage matching mirrors the two stages of heavy-hadron production, i.e. production of the fragmenting heavy-flavor quark at scales between $\sqrt{s}$ and $m_Q$, followed by its hadronization to a heavy-flavor hadron at $\Lambda_{\rm QCD}$ with light degrees of freedom formed around the static heavy source. With the partonic detectors at $\sqrt{s}$ already factorized according to Eq.~\eqref{eq:factorization}, the remaining step matches the first moments of heavy-flavor fragmentation functions onto HQET matrix elements describing the hadronization process with perturbatively computable coefficients~\cite{Mele:1990cw,Neubert:2007je,vonKuk:2023jfd},
\begin{align}
\label{eq:HQET_matching}
D_{i\to H_Q}(N{=}1,\mu) = d^{\rm pert}_{i\to Q}(N{=}1,\mu)\, \chi_Q^{H_Q} + \mathcal{O}\!\left(\frac{\Lambda_{\rm QCD}}{m_Q}\right)\,,
\end{align}
where the HQET matrix element is given by~\cite{Fickinger:2016rfd}
\begin{align}
\label{eq:chi_def}
\chi_Q^{H_Q} &= \frac{1}{4N_c}\sum_{X}{\rm Tr}\,\langle 0|W^\dagger Y_v h_v^{(Q)}|H_Q X\rangle\nonumber\\
&\qquad\times\langle H_Q X|\bar{h}_v^{(Q)} Y_v^\dagger W|0\rangle\,,
\end{align}
with HQET-normalized states and the trace over color and Dirac indices. Here $h_v^{(Q)}$ is the sterile static field of the heavy quark $Q$ with velocity $v$, decoupled from the light degrees of freedom~\cite{Korchemsky:1991zp,Bauer:2001yt,vonKuk:2023jfd}, which carries the heavy-quark degree of freedom at the origin, with its soft color interactions encoded in the timelike Wilson line $Y_v$. The heavy flavor is a label defining a sector, carried by the static source whose interactions with the light degrees of freedom occur only through its fundamental color charge, encoded in $Y_v$. The nonperturbative dynamics at the scale $\Lambda_{\rm QCD}$ resides entirely in the matrix elements of the timelike Wilson line $Y_v$, the static color source of the heavy quark, and the lightlike Wilson line $W$ along the direction $\bar{n}$ opposite to the hadron, inherited from the gauge-invariant definition of the fragmentation function.
Conservation of heavy-quark number within this sector implies the completeness relation
\begin{align}
\label{eq:chi_sumrule}
\sum_{H_Q} f_Q(H_Q)\,\chi_Q^{H_Q} = \sum_{H_Q} \chi_Q^{H_Q} = 1\,, \qquad \chi^{\bar{H}_Q}_{\bar{Q}} = \chi^{H_Q}_{Q}\,,
\end{align}
where the sum runs over a complete set of heavy-flavor states. The flavor weight drops out because the light degrees of freedom dressing the Wilson lines carry no heavy flavor, so every species the source can hadronize into inherits the flavor of the sector defined by $h_v^{(Q)}$, $f_Q(H_Q)=+1$, and the remaining sum expresses that it hadronizes into some species with unit probability. Equivalently, the weighted sum in \eq{chi_sumrule} is the matrix element of the flavor detector itself,
\begin{align}
\label{eq:transparent}
\sum_{H_Q} &f_Q(H_Q)\,\chi_Q^{H_Q} \nn \\
&= \frac{1}{4N_c}{\rm Tr}\,\langle 0|W^\dagger Y_v h_v^{(Q)}\,\mathcal{Q}^{\Lambda}_{f_Q} \bar{h}_v^{(Q)} Y_v^\dagger W|0\rangle = 1\,,
\end{align}
where $\mathcal{Q}^{\Lambda}_{f_Q}$ is the low-energy theory flavor detector restricted to the scale below $m_Q$, along the direction inherited from the matching. At leading power all heavy flavor exits along the source, so the angular delta function factors out and $\mathcal{Q}^{\Lambda}_{f_Q}$ acts as the conserved heavy-flavor charge of the low energy theory. As the light degrees of freedom carry none, the detector is transparent. In terms of \eq{chi_def}, summing over species replaces the projector onto $H_Q$ by the identity, the sterile field $h_v^{(Q)}$ contributes only its normalization, and the Wilson lines cancel by unitarity, $W^\dagger Y_v Y_v^\dagger W=1$. Inserting \eqs{HQET_matching}{chi_sumrule} into \eq{moment_def}, all species-dependent non-perturbative matrix elements cancel, giving
\begin{align}
\label{eq:heavy_sumrule}
\hspace{-0.35cm}\mathcal{M}_q^{f_Q} = d^{\rm pert}_{q\to Q}(1) - d^{\rm pert}_{q\to \bar{Q}}(1)= f_Q(q) + \mathcal{O}\!\left(\frac{\Lambda_{\rm QCD}}{m_Q}\right)\,,
\end{align}
where we also used the fact that the non-singlet difference is fixed to all orders by perturbative heavy-flavor-number conservation. The naive sum rule of \eq{sumrule}, violated non-perturbatively for the light flavors, is therefore satisfied for heavy flavors, up to $\mathcal{O}(\Lambda_{\rm QCD}/m_Q)$ corrections.

\begin{figure}[t]
\centering
\includegraphics[width=0.44\textwidth]{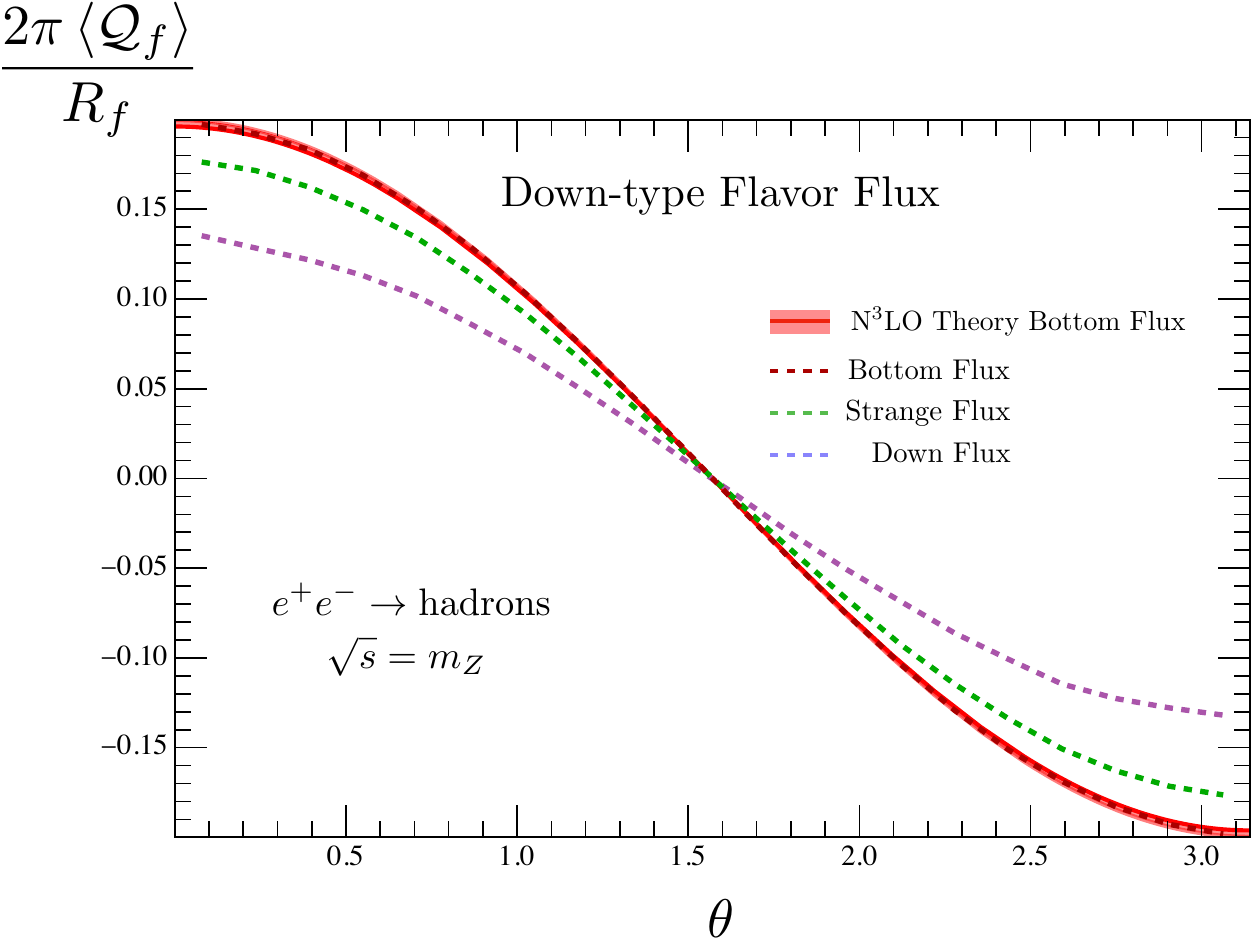}\\
\includegraphics[width=0.44\textwidth]{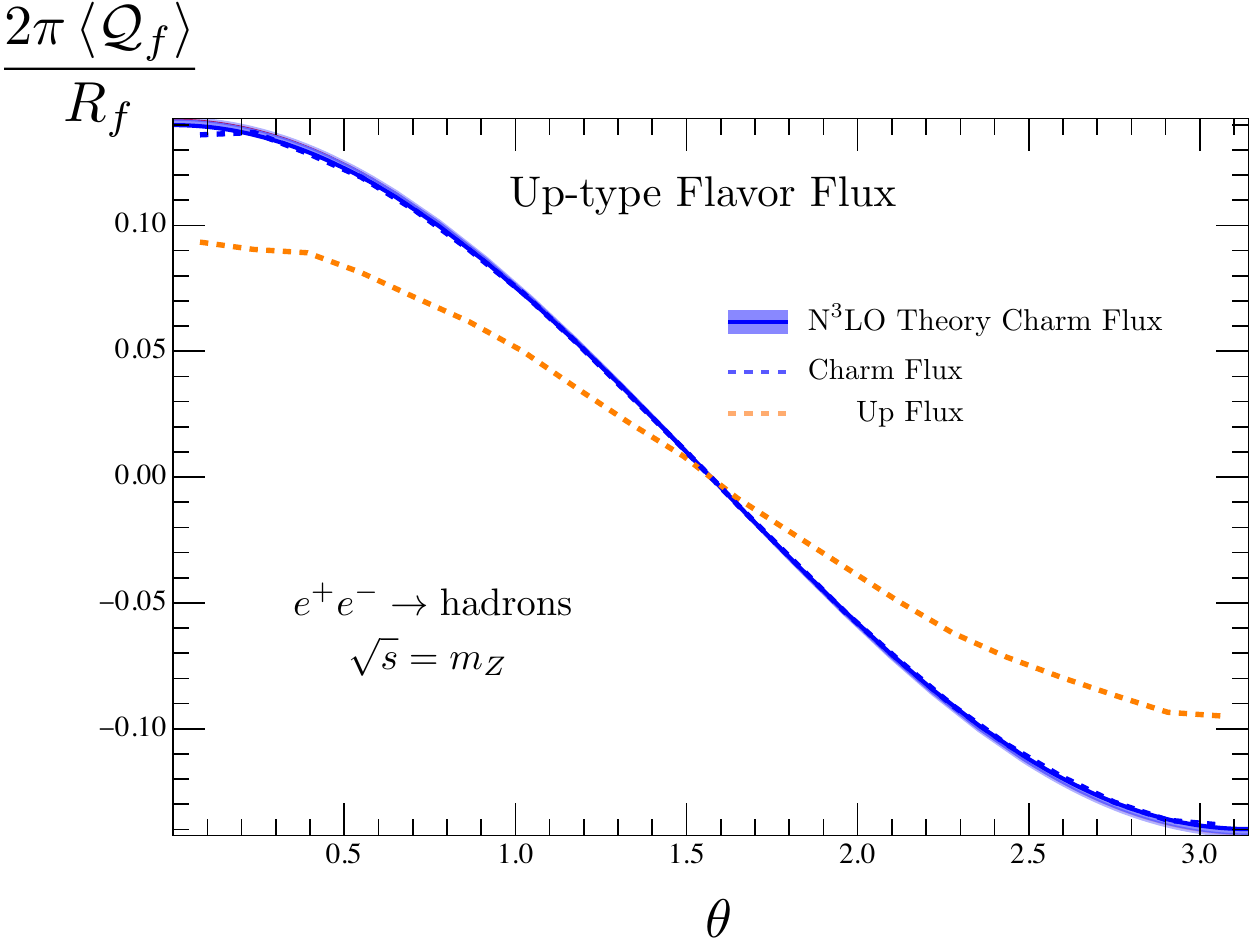}
\caption{Flavor fluxes of down-type (top) and up-type (bottom) flavors in \textsc{Pythia} (dashed), each normalized to events with the corresponding flavor produced at the hard vertex, compared with the prediction of \eq{factorization}, which is identical for all flavors within each class (solid bands, labeled by the heavy flavor of each class). The bottom and charm fluxes saturate the flavor sum rule, while progressively lighter flavors exhibit progressively larger nonperturbative deficits, indicating the loss of flavor to soft hadrons at wide angles.}
\label{fig:FlavorFlow}
\end{figure}

It is instructive to contrast other quantum numbers one might flow through the same machinery. The \emph{electric charge} of heavy hadrons is not protected. Denoting by $e_h$ the electric charge of hadron $h$ in units of the positron charge, consider the electric-charge flux restricted to open heavy-flavor hadrons, $\mathcal{Q}_{\gamma_Q}$, defined by the weight $\gamma_Q(h)=e_h$ for $h\in\{H_Q,\bar{H}_Q\}$ and zero otherwise. Its moment is
\begin{align}
\label{eq:em_moment}
\mathcal{M}^{\gamma_Q}_q \equiv \sum_{H_Q} e_{H_Q}\, \big[D_{q\to H_Q}(N{=}1)-D_{\bar q\to H_Q}(N{=}1)\big]\,,
\end{align}
where we used $e_{\bar{H}_Q}=-e_{H_Q}$. The same HQET matching as in \eq{HQET_matching}, combined with the perturbative sum rule of \eq{heavy_sumrule}, yields
\begin{align}
\label{eq:ebar}
\mathcal{M}^{\gamma_Q}_q = \overline{e}_Q\, f_Q(q) + \mathcal{O}\!\left(\frac{\Lambda_{\rm QCD}}{m_Q}\right)\,,\quad \overline{e}_Q \equiv \sum_{H_Q} e_{H_Q}\,\chi_Q^{H_Q}\,,
\end{align}
where $\overline{e}_Q$, the hadronization-averaged charge of the hadron containing $Q$, is a nonperturbative charge-weighted HQET matrix element that does not have a simple sum rule, as electric charge is also carried by the ungapped light degrees of freedom visible to the Wilson-line dynamics below $m_Q$. The sensitivity to how soft radiation dresses the heavy quark thus reappears, and hadronization rescales the charge flow by the overall factor $\overline{e}_Q/e_Q$.

\section{Reformulating Forward-Backward Asymmetries}
\label{sec:afb}

While we believe that flavor flux correlators will have numerous collider physics applications, here we emphasize their utility with one particularly important application: heavy flavor asymmetries at the $Z$-pole. Forward-backward asymmetries have a long and important history in our understanding of the Standard Model. They can be performed inclusively on all hadrons \cite{OPAL:1997tsq,OPAL:1992jsm,ALEPH:1991fba,ALEPH:1996qlh,ALEPH:1998pmr,L3:1991gfs,L3:1998jgx,DELPHI:1991mqi}, on $b$- and $c$-enriched decays \cite{OPAL:1993wua,ALEPH:2001mdb,L3:1992fsb,DELPHI:2004wvq}, and even on $s$-enriched decays \cite{DELPHI:1994aml,SLD:2000jop,DELPHI:1999mkl}. As discussed in the introduction, when formulated as one-point correlators of charge flux, such asymmetries are governed by anomalies \cite{Hofman:2008ar,Zhang:2026emt}. Here we extend this reformulation to heavy flavor fluxes.

The key advantage of directly reformulating measurements in terms of detector operators, such as \eqref{eq:detector}, is that it allows them to be related to correlation functions of operators. As explained in detail in \cite{Zhang:2026emt}, one-point correlators of charge fluxes, as in \eq{correlator_defN}, are related to three-point functions of currents, in this case $\langle J_Z J_b J_Z \rangle$. Famously, in conformal field theories, three-point functions of currents which exhibit an anomaly are completely fixed by the anomaly coefficient \cite{Schreier:1971um,Osborn:1993cr}. This leads to a characteristic form of the one-point correlator of any charge detector
\cite{Hofman:2008ar}
\begin{align}
\langle \mathcal{Q}(\vec{n})\rangle_{J}&
= \frac{3}{\pi}\frac{d_{J\! J Q}}{c_{J}}\frac{i\epsilon^{ijk} \epsilon^*_i \epsilon_j n_k}{\epsilon^*_l \epsilon_l}\,, 
\label{eq:oneptCharge}
\end{align}
where $\epsilon$ is the polarization vector of the current. In the lab frame, this corresponds to the $\cos(\theta)$ angular asymmetry of the forward-backward asymmetry, and rephrases forward-backward asymmetries at the $Z$-pole as macroscopic manifestations of mixed-electroweak anomalies. 

\begin{figure}[t]
\centering
\includegraphics[width=0.44\textwidth]{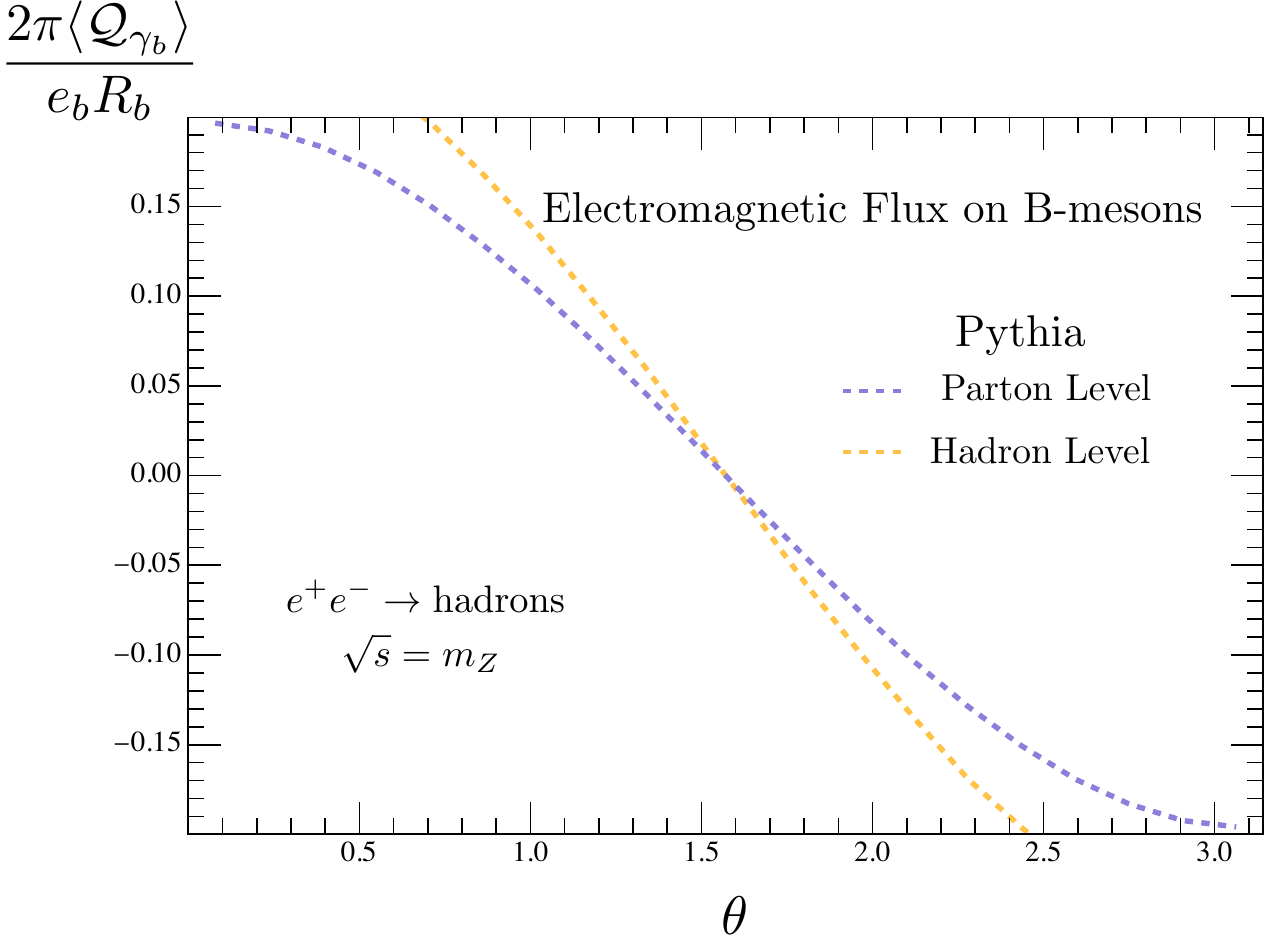}\\
\includegraphics[width=0.44\textwidth]{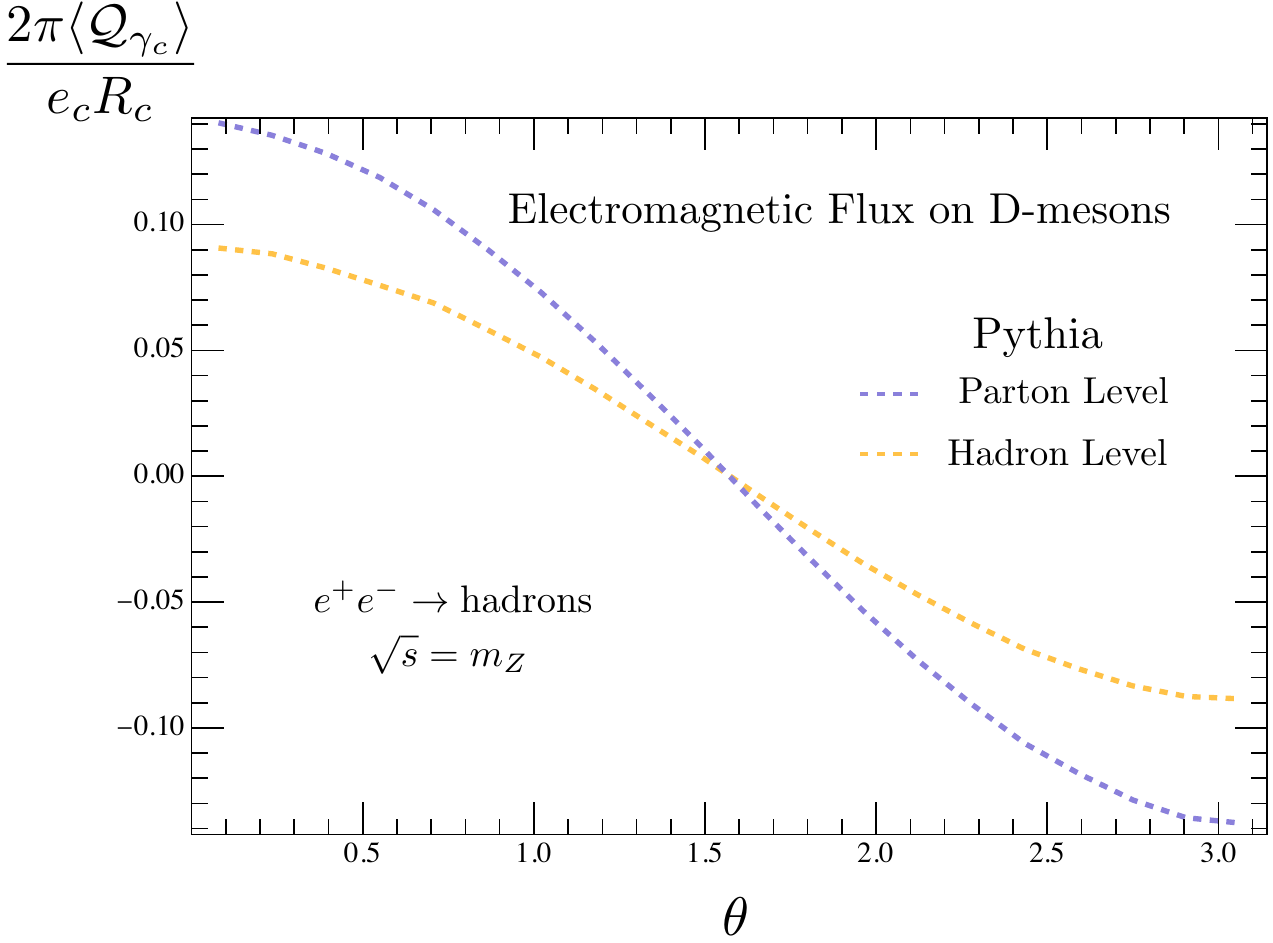}
\caption{Electric charge fluxes carried by bottom (top) and charm (bottom) hadrons in \textsc{Pythia}, $2\pi\langle\mathcal{Q}_{\gamma_Q}\rangle/(e_Q R_Q)$, at parton level (purple) and hadron level (gold). With the $1/e_Q$ normalization, the naive sum rule would reproduce the flavor fluxes of \fig{FlavorFlow}, which the parton-level curves do. Hadronization rescales the charge fluxes by $\overline{e}_Q/e_Q$ of \eq{ebar}, enhancing the bottom and suppressing the charm flux.}
\label{fig:ChargeFlow}
\end{figure}

Using flavor flux correlators, we can rephrase the heavy-quark forward-backward asymmetry in the same manner. As shown in \cite{Zhang:2026emt}, this extends to the case of non-conformal theories, with computable corrections proportional to the $\beta$ function. The $\mathcal{O}(\alpha_s)$ correction to $C_{A}^{\rm NS}$ vanishes identically, with the first correction appearing at $\mathcal{O}(\alpha_s^2)$~\cite{Rijken:1996npa},
\begin{align}
\label{eq:CANS}
C_{A}^{\rm NS} = 1 - \left(\frac{\alpha_s}{4\pi}\right)^2 12\, \beta_0\, C_F\, \zeta_3 + \mathcal{O}(\alpha_s^3)\,,
\end{align}
with $\beta_0 = \tfrac{11}{3}C_A - \tfrac43 T_F n_f$, giving a small correction of $\mathcal{O}(1\%)$ at $\sqrt{s}=m_Z$. In the numerical results below we also include the $\mathcal{O}(\alpha_s^3)$ correction~\cite{Rijken:1996npa}, whose explicit analytic form we do not present here.

As emphasized above, for the case of heavy flavor, the anomaly appears as a matching coefficient, with non-perturbative corrections suppressed by the gapped nature of the flavor quantum number. This is illustrated in \Fig{fig:flavorflow}. Therefore, in the heavy flavor case, one obtains a sharp probe of this physics, as well as a clean reformulation of the forward-backward asymmetry. As compared with previous formulations, it is experimentally much simpler, since it does not make reference to jets or hemispheres. On the theoretical side, the relation to the anomaly offers protection from large perturbative corrections, and the use of flavor as a quantum number protects it from non-perturbative corrections.

To illustrate the picture developed above, we study flavor fluxes in \textsc{Pythia}~8.3~\cite{Bierlich:2022pfr} simulations of $e^+e^-\to Z/\gamma^*\to q\bar{q}$ at $\sqrt{s}=m_Z$. Weak decays are turned off, corresponding to the assumption that heavy mesons can be fully reconstructed. With modern techniques \cite{Keck:2018lcd,Abumusabh:2025tti,Belle-II:2021zvj,Liang:2023wpt,Zhu:2026dko}, we believe this should be possible. For a recent study in the context of forward-backward asymmetries, see \cite{Rohrig:2025bea}. Additional details are discussed in the End Matter. The flux is binned in the angle $\theta$ to the electron beam. For each flavor $f$ we normalize to the sample of events in which the flavor $f$ was produced at the hard vertex, i.e.\ we plot $2\pi\langle\mathcal{Q}_{f}(\vec{n})\rangle/R_f$, and compare with the corresponding prediction \eq{factorization}. At the Born level this normalization makes the contact with the forward-backward asymmetry explicit, $\Afb{f} = \int d \cos \theta\;{\rm sgn}(\cos \theta)\, \pi\langle\mathcal{Q}_f(\vec{n})\rangle/R_f$. Beyond the Born level, the flavor flux correlator is a hadron-level observable in its own right.

The results are shown in \fig{FlavorFlow}, for both up-type and down-type quarks. In red and blue, we show the results of the theoretical calculation in the heavy quark limit. We see that it agrees nearly perfectly with the results from \textsc{Pythia} for $b$ and $c$ type quarks, for which the heavy quark approximation is valid. We emphasize that the \textsc{Pythia} result includes hadronization, so this level of agreement is quite remarkable.

We can emphasize the protection from the $b$ and $c$ quark quantum numbers being gapped in two ways. First, in \fig{FlavorFlow}, we also show \textsc{Pythia} results for light-quark flavor fluxes. The perturbative calculations of the light-quark flavor fluxes are identical. However, as discussed above, since these quantum numbers are not gapped at a scale above $\Lambda_{\text{QCD}}$, they receive non-perturbative corrections, which should scale like $e^{-\#\, m_Q^2/\Lambda_{\rm QCD}^2}$, i.e. the naive sum rule of \eq{sumrule} is violated at $\mathcal{O}(1)$ for the light flavors. The rapid increase in non-perturbative corrections as the quark mass is lowered is clearly seen in \fig{FlavorFlow}, and emphasizes the uniqueness of heavy flavor flux. As emphasized in \cite{Ke:2026zia}, these non-perturbative corrections are interesting probes of hadronization, and it would be interesting to understand them better. 

A second way of emphasizing the importance of the use of flavor flux is to compare with the flux of electromagnetic charge restricted to B-mesons. Indeed, in previous measurements of forward-backward asymmetries, charge was often used to associate hemispheres with $b$ or $\bar b$ mesons. The electric charge fluxes carried by the bottom and charm hadrons, $2\pi\langle\mathcal{Q}_{\gamma_Q}\rangle/(e_Q R_Q)$ for $Q=b,c$, are shown in \fig{ChargeFlow}, normalized by the quark charge $e_Q$ so that the naive sum rule, $\mathcal{M}_q^{\gamma_Q}\to e_Q\, f_Q(q)$, would place them exactly on top of the flavor fluxes of \fig{FlavorFlow}. At parton level we see that this is indeed true, and the charge fluxes coincide with the bottom and charm flavor fluxes of \fig{FlavorFlow}. However, since electromagnetic charge is not gapped, the charge fluxes are rescaled at hadron level by $\overline{e}_Q/e_Q$ of \eq{ebar}. Predicting them therefore requires the nonperturbative input $\overline{e}_Q$, in contrast to the flavor flux, which remains a zero-parameter prediction. This emphasizes the unique robustness of gapped flavor fluxes.

\section{Conclusions}
\label{sec:conclusions}

Motivated by the essential role flavor plays in collider physics experiments, in this \emph{Letter} we introduced flavor flux correlators, incorporating flavor into the energy correlator paradigm. We pointed out that, as compared to other charges, heavy flavor fluxes are particularly clean, since heavy flavor quantum numbers are gapped at a heavy quark mass scale $m_Q \gg \Lambda_\text{QCD}$, enabling them to be entirely computed in perturbation theory, up to $\mathcal{O}(\Lambda_{\rm QCD}/m_Q)$ corrections. We illustrated this robustness for the one-point correlator in Pythia, and contrasted it to electromagnetic charge, which is gapped at the confinement scale and receives much larger non-perturbative corrections.

As an illustration of the utility of flavor flux correlators, we reformulated one of the key Standard Model measurements, namely heavy quark forward-backward asymmetries at the $Z$-pole, as a one-point correlator of heavy flavor flux. Due to the operator definition of flavor flux correlators, this allows us to relate this foundational measurement to the three-point function of currents, $\langle J_Z J_b J_Z \rangle$, and ultimately to a mixed electroweak anomaly of the b-flavor current in the Standard Model. This formulation is under precise theoretical control, with perturbative corrections starting at $\mathcal{O}(\alpha_s^2)$ and non-perturbative corrections at $\mathcal{O}(\Lambda_{\rm QCD}/m_Q)$, and it avoids experimental ambiguities associated with jet or hemisphere definitions.

From the experimental perspective, measurements of flavor flux correlators should be possible at multiple experiments. Archival LEP and SLC data, combined with modern flavor tagging~\cite{Defranchis:2026wyw}, would enable a first measurement of bottom and charm fluxes on existing $Z$-pole samples. We can also measure flavor flux off the $Z$ pole, and Belle~II's dedicated charm-tagging capabilities~\cite{Belle-II:2023vra} make it a natural target there. At a future Tera-$Z$ facility~\cite{FCC:2018evy,CEPCStudyGroup:2018ghi,Gori:2015nqa,Allwicher:2025bub}, the enormous $Z$ samples would push the flux-correlator measurements of the heavy-flavor asymmetries to much higher precision. In all cases, a detailed understanding of tagging efficiencies for different decay modes, as well as a treatment of neutral meson oscillations, will be required. A brief discussion of experimental issues is presented in the End Matter. 

On the theory side, several directions merit further development, including heavy-quark mass power corrections to \eq{CANS} and a universal operator definition of the non-perturbative violations of the flavor sum rule across quark species. This would further sharpen the theoretical control of flavor flux observables and extend their reach beyond the heavy-quark limit. It would also be interesting to explore the applications of flavor flux correlators to other collider systems, such as electron-proton collisions, or to extend recent proposals of charge flux measurements \cite{Cao:2026fzq} or nucleon-energy correlators \cite{Liu:2022wop}, by adding flavor.

More broadly, our work is part of a program to reformulate foundational electroweak measurements in the language of asymptotic detector observables. This framework brings the exceptional theoretical control of detector correlators directly to precision tests of the Standard Model, opening a new avenue for revisiting longstanding tensions with existing and future collider data.

\emph{Acknowledgements.}---We thank Justin Kulp, Yen-Jie Lee, Zoltan Ligeti, Xiaohui Liu, Michael Peskin, Huilin Qu, Manqi Ruan, and Jingyu Zhang for useful discussions.
K.L. is supported by the U.S. Department of Energy under contract DE-AC02-06CH11357.
 I.M. is supported by the DOE Early Career Award DE-SC0025581, the Sloan Foundation, and the Simons Collaboration on Confinement and QCD Strings.
\section{End Matter}
\subsection{Practical Experimental Considerations}
\label{sec:mixingandtagging}

The discussion in the main text assumed an idealized scenario of perfect identification of the flavor of the heavy hadrons. In practice, two issues must be addressed: the heavy hadrons must be tagged and reconstructed with realistic efficiencies, and neutral mesons oscillate before decaying. Both effects enter the flux in the same way, as modifications of the completeness relation \eq{chi_sumrule}, and both are brought under control by precision knowledge of the individual HQET matrix elements $\chi_Q^{H_Q}$.

Consider first imperfect tagging. The sum rule \eq{chi_sumrule} implicitly assumed that every open heavy-flavor species is identified with $100\%$ efficiency. Denoting by $\upepsilon(H_Q)$ the efficiency for tagging and reconstructing the species $H_Q$, a realistic measurement is instead sensitive to
\begin{align}
\label{eq:eff_sumrule}
\sum_{H_Q} f_Q(H_Q)\, \chi_Q^{H_Q}\, \upepsilon(H_Q)\,,
\end{align}
which no longer sums to unity. The completeness relation is broken not by QCD but by the detector, and the individual $\chi_Q^{H_Q}$ no longer cancel. Note that the angular structure of \eq{factorization} is untouched, since the $\upepsilon(H_Q)$ are properties of the detector response to identified hadrons. Therefore, the entire effect is a rescaling of the normalization.

The second issue arises because hadronization produces flavor eigenstates, while the states of definite mass and lifetime are superpositions of the neutral mesons with their antiparticle counterparts. The $\bar{B}^0$ and $\bar{B}_s^0$ therefore oscillate before they decay, driven by the weak interaction, and a flavor tag built on the decay products returns the production flavor only up to this oscillation. Denoting by $g^{\rm mix}_q$ the time-integrated probability for the neutral meson to decay as its antiparticle (denoted $\chi_q$ in the literature, a symbol we reserve for the HQET matrix elements), each neutral species analyzes its production flavor with dilution $1-2g^{\rm mix}_q$. A decay-tagged measurement is then sensitive to \eq{eff_sumrule} further modified as
\begin{align}
\label{eq:tagged_moment}
\sum_{H_b} f_b(H_b)\, \chi_b^{H_b}\,\upepsilon(H_b) &- 2\, g^{\rm mix}_d\, \chi_b^{\bar{B}^0}\,\upepsilon(\bar{B}^0)\nonumber\\
&- 2\, g^{\rm mix}_s\, \chi_b^{\bar{B}_s^0}\,\upepsilon(\bar{B}^0_s)\,,
\end{align}
with current world averages $g^{\rm mix}_d = 0.1860 \pm 0.0011$ and $g^{\rm mix}_s = 0.499314\pm 0.000005$~\cite{HeavyFlavorAveragingGroupHFLAV:2024ctg}. The $\bar{B}^0$ tag is thus partially diluted, $1-2g^{\rm mix}_d \simeq 0.63$, while for the $\bar{B}_s^0$ the rapid oscillation almost completely erases its contribution to the signed flux, $1-2g^{\rm mix}_s \simeq 1.4\times 10^{-3}$. Mixing was indeed a nontrivial source of systematic uncertainty in the previous lepton-tag measurements of $\Afb{b}$~\cite{OPAL:2003pfe,DELPHI:2003fml}. For charm, $D^0$ mixing is minuscule, $g^{\rm mix}_D \sim 10^{-5}$, so the charm flux is effectively undiluted.

The structure of \eqs{eff_sumrule}{tagged_moment} makes clear how these practical difficulties are overcome in principle. The $\cos\theta$ shape of \eq{factorization} is unaffected, and the normalization is modified by three inputs: the detector efficiencies $\upepsilon(H_b)$, the mixing probabilities $g^{\rm mix}_q$, and the HQET matrix elements $\chi_b^{H_b}$. One can then either take the $\chi_b^{H_b}$, $g^{\rm mix}_q$, and $\upepsilon(H_b)$ as inputs and simply compute \eq{tagged_moment}, obtaining an exact prediction for the flux measured on any tagged subset of species. Or one can undo the modification at the level of the measurement by weighting each tagged hadron of species $H_b$ by its tagged flavor sign times $1/[\upepsilon(H_b)\,\mathcal{D}_{H_b}]$, with $\mathcal{D}_{H_b} = 1-2g^{\rm mix}_q$ for the neutral species $\bar{B}^0, \bar{B}_s^0$ and $\mathcal{D}_{H_b}=1$ otherwise, which restores the exact sum rule \eq{chi_sumrule} in expectation. The first strategy takes as its only nonperturbative input the finite set of numbers $\chi_b^{H_b}$, which are scale-invariant, universal matrix elements constrained by the completeness relation \eq{chi_sumrule} and measurable species by species, together with the precisely measured $g^{\rm mix}_q$ and the detector-calibrated $\upepsilon(H_b)$. The second requires no knowledge of the $\chi_b^{H_b}$ at all, since the sum rule is restored by construction. The price is that the weights $1/[\upepsilon(H_b)\,\mathcal{D}_{H_b}]$ blow up for poorly tagged or strongly diluted species, and a species not tagged at all, $\upepsilon(H_b)=0$, can only be restored by supplying its $\chi_b^{H_b}$ as an external input. Either way, hadronization enters only through a few well-defined nonperturbative constants, rather than through modeling of its dynamics.

Concerning the experimental inputs themselves, measuring $\langle\mathcal{Q}_{f_Q}(\vec{n})\rangle$ requires, event by event, identified heavy hadrons with (i) their direction and (ii) the sign of their heavy-flavor quantum number. Neither the hadron energy nor a complete reconstruction of the event is needed. This is a substantially weaker requirement than reconstructing fragmentation spectra. At LEP and SLC, heavy-flavor hadrons were tagged through their semileptonic decays~\cite{OPAL:2003pfe,DELPHI:2003fml,L3:1992fsb}, through $D^{*\pm}$ reconstruction~\cite{OPAL:1993wua}, and through lifetime tags with the flavor sign estimated from jet and vertex charges~\cite{ALEPH:1996qlh,ALEPH:1998pmr,ALEPH:2001mdb,DELPHI:2004wzo}. In recent years, machine-learned flavor tagging has dramatically improved hadron- and jet-level identification of $b$, $c$, and even $s$ quarks, and has been deployed on archival ALEPH data~\cite{Defranchis:2026wyw}, demonstrating that modern taggers can be run on the existing $Z$-pole samples. In the context of FCC-ee, \Ref{Rohrig:2025bea} has shown that \emph{exclusive} $b$-hadron reconstruction over an $\mathcal{O}(200)$-mode list achieves flavor tags of purity above $99.7\%$ at efficiencies of order one percent. For charm, exclusive $D^{(*)}$ reconstruction is standard, and Belle~II has recently commissioned a dedicated flavor tagger for the production flavor of neutral $D$ mesons~\cite{Belle-II:2023vra}.

\begin{figure}[t]
\centering
\includegraphics[width=0.48\textwidth]{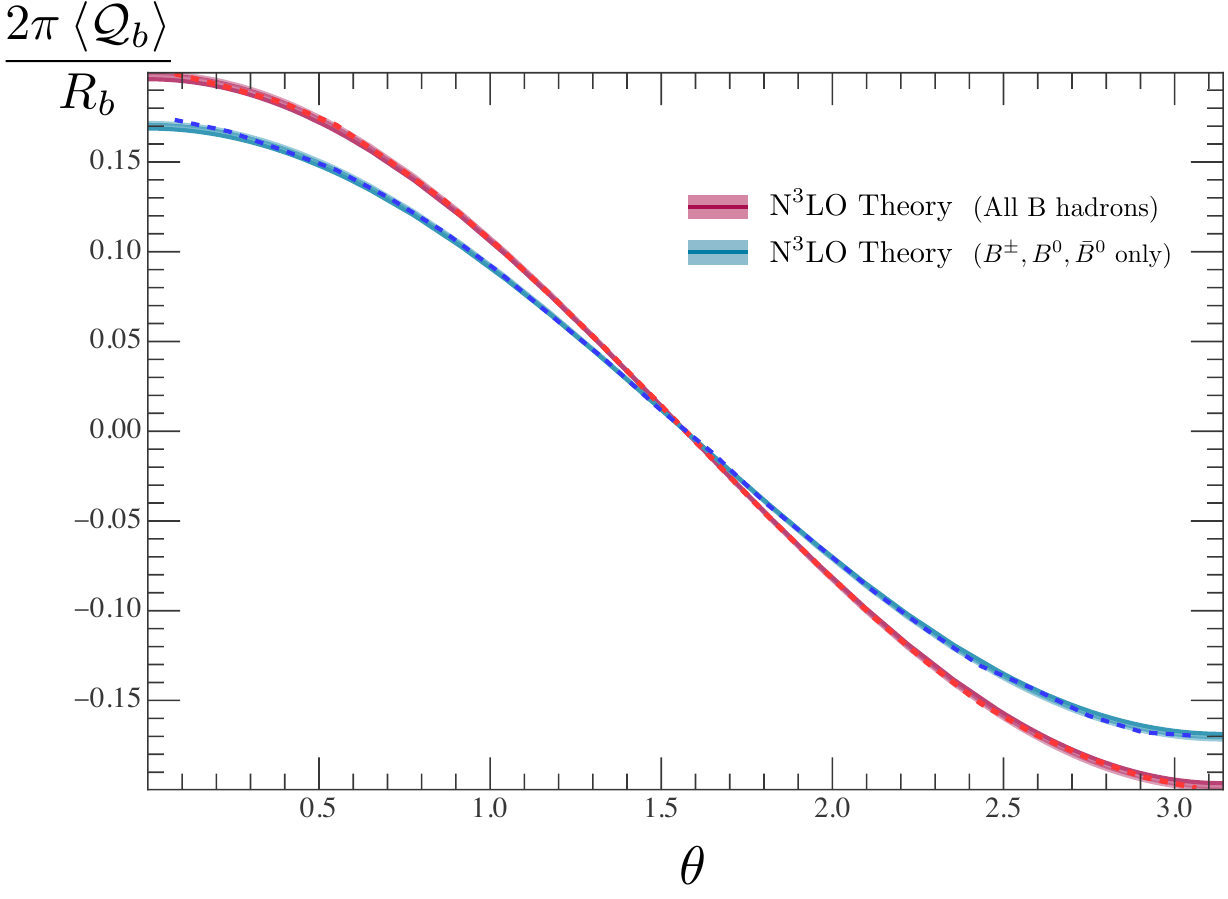}
\caption{The bottom-flavor flux in \textsc{Pythia} (black dashed) compared with the prediction of \eq{factorization} (solid bands), measured on all $b$ hadrons (red) and restricted to the tagged species $B^\pm, B^0, \bar{B}^0$ only (blue). }
\label{fig:HadronTag}
\end{figure}

As a simple demonstration, consider the scenario $\upepsilon(H_b) = 1$ for $H_b \in \{B^\pm, B^0, \bar{B}^0\}$ and $\upepsilon(H_b)=0$ for all other $b$ hadrons. Since the Monte Carlo measurement above is performed at the production level, mixing does not enter, and \eq{tagged_moment} reduces to $\chi_b^{B^-}+\chi_b^{\bar{B}^0}$. Isospin symmetry equates the two matrix elements, and with $\chi_b^{B^-}=\chi_b^{\bar{B}^0} = 0.407\pm 0.007$~\cite{HFLAV:2022esi} the restricted flux is predicted to be $\simeq 0.814\pm 0.014$ of the full one, with no change in shape. In \textsc{Pythia} the restricted flux indeed retains the shape of the full one, with a normalization $\simeq 0.86$, the value used for the band in \fig{HadronTag}, very close to though slightly above the data-driven expectation. The small excess stems from \textsc{Pythia} producing slightly larger $B^\pm, B^0, \bar{B}^0$ fractions than the measured values. The same exercise with decay-level tagging would in addition carry the $1-2g^{\rm mix}_q$ dilution of the neutral species, computable in exactly the same way. This illustrates the general lesson of this section. Once the $\chi_b^{H_b}$ are known, the flavor flux remains a zero-parameter prediction for any realistic tagging configuration.

\bibliography{refs}

\end{document}